\documentclass[12pt]{article}

\usepackage{amssymb}
\usepackage{epsfig}
\newcommand{\labl}[1]{\label{#1}}

\newcommand{\eq}{\begin{equation}}
\newcommand{\eqx}{\end{equation}}
\newcommand{\eqn}{\begin{eqnarray}}
\newcommand{\eqnx}{\end{eqnarray}}
\newcommand{\f}[2]{\frac{#1}{#2}}

\def\al{\alpha}
\def\la{\lambda}
\def\bx{\bar{x}}

\def\e {{\rm e\,}}
\newcommand\widebar[1]{\overline{\; #1}}

\begin{document}

\begin{titlepage}
\begin{flushright}
\begin{tabular}{l}
LPT--Orsay--99--54\\ TPJU-6/99 \\ hep-ph/9908304
\end{tabular}
\end{flushright}

\vskip3cm

\begin{center}
  {\large \bf
   New representation for the odderon wave function
  \\}

\vspace{1cm}
{\sc G.P.~Korchemsky}${}^1$ and {\sc J. Wosiek}${}^2$
\\[0.5cm]

\vspace*{0.1cm} ${}^1$ {\it
Laboratoire de Physique Th\'eorique%
\def\thefootnote{\fnsymbol{footnote}}%
\footnote{Unite Mixte de Recherche du CNRS (UMR 8627)},
Universit\'e de Paris XI, \\
91405 Orsay C\'edex, France
                       } \\[0.2cm]
\vspace*{0.1cm} ${}^2$ {\it
Institute of Physics, Jagellonian University \\
Reymonta 4, 30-059 Cracow, Poland
                       } \\[1.0cm]

\end{center}

\begin{abstract}
New representation of the odderon wave function is derived, which
is convergent in the whole impact parameter plane, and  provides the
analytic form of the quantization condition for the integral of motion $q_3$.
A new quantum number, triality, was identified independently in each sector.
This, together with the choice of the conformal basis allows for simple calculation
of eigenvalues of a wide class of operators.
\end{abstract}

\end{titlepage}

\newpage
\setcounter{page}{2}
\section{Introduction}
The resummed scattering amplitudes in perturbative QCD grow with
the energy in a qualitative agreement with the Regge theory
expectations. Till recently most of the activity has been concentrated
around the leading Regge trajectory with the vacuum quantum
numbers -- the BFKL pomeron \cite{BFKL}. 
%At the same time,
Perturbative QCD predicts also an existence of the object with the
same quantum numbers except of the C parity -- the odderon
\cite{firstb,LipOd,GLN}, which controls the energy behavior of the
difference of the cross-sections, $\sigma^{\rm tot}_{pp} -
\sigma^{\rm tot}_{p\bar p}\sim s^{\alpha_O-1}$, with $\alpha_O$
being the odderon intercept. Contrary to the BFKL pomeron, the
intercept of the odderon trajectory has been calculated only
recently \cite{USPRL,PRL2} which generated renewed interest in the
subject \cite{Brau,BGN,LIPRP}. In this letter we present an
alternative derivation of the odderon wave function. Present
approach leads to simpler (although equivalent) final expression.
%Also 
In addition, it offers a possibility of a more direct calculation of the
odderon intercept.

In the perturbative QCD approach, the odderon appears as a compound
state of three gluons bound together by a nontrivial QCD interaction \cite{firstb}.
The odderon wave function satisfies the Schr\"odinger equation with
the effective QCD Hamiltonian $\mathbb{H}_3$ describing quantum mechanical
3-body system with nearest neighbor interaction on a two-dimensional
plane of transverse gluon coordinates $\vec\rho=(x,y)$.
The odderon intercept, $\alpha_O-1$,
is given by the maximal eigenvalue of the
Hamiltonian and the corresponding eigenfunction determines the odderon
wave function $\chi_O(\vec\rho_1,\vec\rho_2,\vec\rho_3)$.

The effective QCD Hamiltonian, $\mathbb{H}_3$, possesses a number of
remarkable properties 
%that 
which allow to solve the odderon problem
exactly \cite{LipOd,FK}. Firstly, it exhibits the holomorphic separability
\begin{equation}
\mathbb{H}_3= {\cal H}_3 + \widebar {\cal H}_3\,,\qquad
 [{\cal H}_3,\widebar {\cal H}_3]=0,
\label{H-QCD}
\end{equation}
with the operators ${\cal H}_3= H_{12} + H_{23} + H_{31}$ and
$\bar {\cal H}_3$ acting on
the holomorphic and antiholomorphic gluon coordinates,
$\rho_k=x_k+i y_k,$ and $\bar{\rho}_k=x_k-i y_k$
$(k=1,2,3)$, respectively. The two-body holomorphic Hamiltonian is
given by
\begin{equation}
H_{ik}=\psi(J_{ik})+\psi(1-J_{ik})-2\psi(1)\,,\qquad
J_{ik}(1-J_{ik})\equiv L^2_{ik}=-\rho_{ik}^2\partial_i\partial_k.
\label{hik}
\end{equation}
Secondly, in each sector two additional operators exist, which
commute with the Hamiltonians ${\cal H}_3$ and $\bar {\cal H}_3$
\begin{equation}
q_2\equiv L^2=L^2_{12}+L^2_{23}+L^2_{31}\,,\qquad
q_3=i \rho_{12}\rho_{23}\rho_{31}\partial_1\partial_2\partial_3,
\end{equation}
with similar expressions for the antiholomorphic operators. The
conserved charge, $q_2$, is given by the quadratic Casimir operator
of the $SL(2)$ group and is related to the invariance of ${\cal
H}_3$ under projective transformations $\rho\to
(a\rho+b)/(c\rho+d)$ with $ad-bc=1$. The existence of the
additional conserved charge $q_3$ leads to a complete
integrability of the QCD Hamiltonian $\mathbb{H}_3$, and plays a
crucial role in our consideration.

Thanks to the projective invariance of the Hamiltonian
$\mathbb{H}_3$, the wave function $\chi(\rho_k,\bar{\rho}_k)$ can
be classified according to the irreducible representation of the
$SL(2,\mathbb{C})$ group. The choice of a particular
representation is dictated by additional physical requirement that
one has to impose on the properties of the odderon state. As in
the case of the BFKL pomeron, one chooses the appropriate
representation as the unitary principal series of the
$SL(2,\mathbb{C})$. It is straightforward to show that under this
choice the eigenvalues of the Hamiltonian $\mathbb{H}_3$
(including the value of the odderon intercept) take {\it finite\/}
real values \cite{FK}. The corresponding eigenfunctions are parameterized by
conformal weights $h={1\over 2}(1+m) -i\nu$ and $\bar h=1-h^*$ and
have the following general form
\begin{equation}
\chi(\rho_k,\bar\rho_k)
= \left({\rho_{12}\over \rho_{10} \rho_{20}} \right)^{h}
  \left({\bar\rho_{12}\over \bar\rho_{10} \bar\rho_{20}} \right)^{\bar h}
\Psi(x,\bar x).
\label{chi}
\end{equation}
Here, $h-\bar h=m$ and $h+\bar h=1-2i\nu$ are integer Lorentz spin and
the scaling dimension of the state, respectively, $\vec\rho_0$ is the
center-of-mass coordinate of the state and $\Psi(x,\bar x)$ is
some function of the anharmonic ratio
$x={\rho_{12}\rho_{30}\over\rho_{10}\rho_{32}}$ with
$\rho_{ij}=\rho_i-\rho_j$ and similarly for $\bar x$.
The Bose symmetry of the odderon state implies that
$\chi(\rho_k,\bar\rho_k)$ has to be a completely symmetric
function of gluon coordinates.

For a given $h$ and $\bar h$ and any $\Psi(x,\bar x)$, the function
(\ref{chi}) diagonalizes the $SL(2,\mathbb{C})$ Casimir operators
$q_2$ and $\bar q_2$ and the corresponding eigenvalues are given
by $q_2=h(1-h)$ and $\widebar q_2=\bar h(1-\bar h)$. The explicit
form of $\Psi(x,\bar x)$ should be found from the condition for
$\chi$ to be an eigenstate of the QCD Hamiltonian (\ref{H-QCD}).
Due to a complete integrability, the eigenproblem for
$\mathbb{H}_3$ becomes equivalent to a simpler condition for the function
(\ref{chi}) to be a simultaneous eigenfunction of the conserved
charges $q_3$ and $\bar{q}_3$. Since these operators act
independently on the holomorphic and antiholomorphic coordinates,
$\Psi(x,\bar x)$ has the following general form
\begin{equation}
\Psi(x,\bar x)=\sum_{\lambda,\widebar\lambda}
f_{\lambda\widebar\lambda}\ \Phi_\lambda(x) \widebar{\Phi}_{\widebar\lambda}
(\bar x),
\label{odfun}
\end{equation}
where sum goes over the eigenfunctions $\Phi_\lambda$
$(\bar{\Phi}_{\bar\lambda})$ of the conserved charge $q_3$ ($\bar
q_3$) in the holomorphic (antiholomorphic) sectors and
$f_{\lambda\widebar\lambda}$ are some numerical coefficients.
Although the dynamics in $x$ and $\bar x-$coordinates is
independent of each other, the two sectors are tied together through
the condition that the wave function $\chi(\rho_k,\bar{\rho}_k)$ is
 a single-valued function on the two-dimensional $\rho-$plane.
It is this condition that allows to establish the quantization of
the conserved charges $q_3$ and $\bar q_3$ and uniquely fix the
expansion coefficients $f_{\lambda\widebar\lambda}$ entering into
(\ref{odfun}). In this way, the exact spectrum of $q_3$ and $\bar
q_3$ was recently derived in \cite{PRL2} and found to agree with
the earlier asymptotic WKB expressions \cite{KorQ}.

In Sect.~3 we derive a new analytical quantization condition of
$q_3$ that agrees with the both approaches. Our approach is based
on the observation that in addition to $q_2$ and $q_3$ there
exists the new quantum number in each sector. Namely the
Hamiltonian (\ref{H-QCD}) is invariant under the cyclic
permutations of the reggeon coordinates independently in the
holomorphic and antiholomorphic sectors. Denoting the generators
of the corresponding transformations as $P$ and $\bar P$, we shall
choose $\Phi_\lambda(x)$ in (\ref{odfun}) to be simultaneous
eigenfunctions of $q_2$ and $q_3$ possessing a definite
``triality" $\lambda$ $(\lambda^3=1)$ with respect to cyclic
permutations $P$. This, together with the conformal basis
introduced in the next section, allows for simple calculation of
 eigenvalues of a wide class of operators including the QCD
Hamiltonian.

\section{Conformal basis and cyclic symmetry}

The requirement for the wave function (\ref{chi}) to diagonalize
the conserved charge $q_3$  leads to the 3rd order ordinary
differential equation of the Fuchs type for $\Psi(x,\bar x)$ as a
function of $x$. It proves convenient to analyze this equation by
expanding $\Psi(x,\bar x)$ over the conformal basis spanned by the
functions $\phi_\alpha(x)\phi_{\bar\alpha}(\bar x)$ which,
in the main channel $(123)$, have the
same conformal weights $h$ and $\bar h$ 
 as the wave function $\Psi$ and, in addition, have
a definite conformal weights $\alpha$ and $\bar\alpha$ in the
$(12)-$subchannel \footnote{Here we introduced the notation for
the ``hatted'' operators $\hat L_{jk}^2$ that act on $x$ and are
related to $L_{jk}^2$ through the similarity transformation, $\hat
L_{jk}^2=U^{-1} L_{jk}^2 U$ with
$U=(\rho_{12}/\rho_{10}\rho_{20})^h$.}
\begin{equation}
\hat L^2 \phi_{\al}(x)=h(h-1)\phi_{\al}(x)\,,\qquad
\hat L^2_{12} \phi_{\al}(x)=\al(\al-1)\phi_{\al}(x).
\label{eia}
\end{equation}
It is straightforward to show that the solution of (\ref{eia}) is
given by a hypergeometric series
\footnote{Similar conformal basis was previously introduced in \cite{LIPRP,BDKM}.}
\begin{equation}
\phi_{\al}(x)=x^{\al-h}F(\al, \al-h; 2\al;x),  \label{harmo}
\end{equation}
and one finds similar expressions in the antiholomorphic sector.
As mentioned in the Introduction, this choice of basis, together
with the cyclic symmetry discussed below, allows a rather simple
representation for the large class of operators.

 %Indeed,
Using the
relation $q_3=-i[L^2_{23},L^2_{31}]$ one verifies that the
conserved charge $\hat q_3=U^{-1} q_3 U$ is given in the conformal
basis by, an infinite-dimensional, three-diagonal matrix
\begin{equation}
 i\hat{q}_3\phi_{\al}(x)= A(\alpha)\phi_{\al-1}(x)
+ B(\alpha)\phi_{\al+1}(x),
\label{q3-mat}
\end{equation}
with $$ A(\alpha)=\al(\al-1)(h-\al)\,,\quad B(\alpha) = -
A(\alpha){(h+\al)(h+\al-1)\over 4(2\al+1)(2\al-1)} . $$
The eigenfunction of the operator $\hat q_3$ is given by a linear
combination of the basis functions that one can effectively
represent in the form of the contour integral (we omit the indices
for simplicity)
\begin{equation}
   \Phi(x)=\int_{\Gamma} {d\al\over 2\pi i}\, C(\al)
   \phi_{\al}(x)\,.
\label{inal}
\end{equation}
Here $C(\al)$ is some coefficient function having (an infinite)
number of poles in the complex $\alpha-$plane and the integration goes
over some contour $\Gamma$ encircling all singularities of
$C(\al)$. If $C(\al)$ has a pole at $\alpha=\alpha_0$, then its
contribution to $\Phi(x)$ scales as $x^{\alpha_0}$ at small $x$.
Consequently, requiring $\Phi(x)$ to be an eigenfunction of $\hat q_3$ and
applying (\ref{q3-mat}), one finds that $C(\al)$ should have two
additional poles situated at $\al=\al_0\pm 1$ for any $\al_0$
except of those satisfying $A(\al_0)=0$, or equivalently
$\al_0=0\,,1\,, h$. Each solution to this equation generates a
string of poles shifted by $1$ along the real axis. The fact, that
the first two strings become degenerate starting from $\al=1$
gives rise to a double pole, or equivalently, generates
logarithmic singularity at $x\sim 0$. An additional degeneracy
occurs when conformal weight of the state $h$ takes integer
positive values. In this case, the third series becomes degenerate
with the first two and gives rise to triple poles. In what follows
we will not consider this case separately. Summarizing,
%the properties of the coefficient function
 we find that for arbitrary values
of $h$ (except of integer values) $C(\al)$ admits the following
expansion
\begin{equation}
C(\al)=\sum_{k=0}^{\infty} {c_k\over \al-k}+{a_k\over(\al-k)^2}+{b_k\over \al-k-h},\;\;\;a_0=0,
   \label{cal}
\end{equation}
and the contour $\Gamma$ in (\ref{inal}) is such that all poles
(and only poles) saturate the integral.

Substituting (\ref{cal}) into (\ref{inal}), and using
(\ref{q3-mat}), one finds that for $\Phi(x)$ to be an eigenfunction of $\hat
q_3$ the residue coefficients $(a_k,b_k,c_k)$ have to satisfy the
following three-term recurrence relations
\begin{eqnarray}
 iq_3 a_k &=& a_{k+1}A(k+1)-a_{k-1}B(k-1)\,, \label{rra} \\[3mm]
 iq_3 b_k &=& b_{k+1}A(k+h+1) - b_{k-1}B(k+h-1)\,,\labl{rrb}
 \\[2mm]
 iq_3 c_k &=& \left(c_{k+1}+a_{k+1}\f{d}{dk}\right)A(k+1)
           - \left(c_{k-1}+a_{k-1}\f{d}{dk} \right)B(k-1),
           \label{rrc}\\[2mm]
  &&\hspace*{-18mm} a_0=b_{-1}=c_{-1}=0. \nonumber
\end{eqnarray}
This system determines all coefficients $a_k,b_k$ and $c_k$ in terms
of the three initial ones, which were chosen as $b_0,c_0$ and $c_1$. A
freedom in choosing the initial conditions corresponds to the existence of the
three linearly independent solutions to the third order differential equation 
for  the
eigenfunctions $\Phi_{\la}(x)$ of the operator $\hat q_3$ in the holomorphic
sector. The general expression for (\ref{inal}) looks like
\begin{equation}
\Phi_{\la}(x)=c_0(\lambda)\phi^a(x)+b_0(\la)\phi^b(x)+c_1(\la)\phi^c(x), \label{filam}
\end{equation}
where $\lambda$ enumerates three different eigenfunctions and the
functions $\phi^a$, $\phi^b$ and $\phi^c$ are defined as three
independent solutions to the recurrence relations
Eqs.(\ref{rra})-(\ref{rrc}) corresponding to three different
choices of the initial conditions, $(c_0=1,b_0=0,c_1=0)$,
$(c_0=0,b_0=1,c_1=0)$ and $(c_0=0,b_0=0,c_1=1)$, respectively
\begin{eqnarray}
\phi^a(x)&=&\sum_{k=0}^{\infty}c_k\phi_k(x)+a_k
\partial_k\phi_k(x),\quad
c_0=1,\ c_1=0,\nonumber \\
\phi^b(x)&=&\sum_{k=0}^{\infty}b_k \phi_{k+h}(x),\quad b_0=1\,,\label{fiabc} \\
\phi^c(x)&=&\sum_{k=0}^{\infty}c_k \phi_k(x),\quad c_0=0,\ c_1=1\,.\nonumber
\end{eqnarray}
To fix uniquely the coefficients $c_0(\lambda)$, $b_0(\lambda)$
and $c_1(\lambda)$ we explore an additional {\it discrete\/} symmetry
of the holomorphic Hamiltonian ${\cal H}_3$ and the charges $q_2$ and
$q_3$ with respect to the cyclic permutations of the holomorphic
gluon coordinates
\begin{equation}
P\, \chi(\rho_1,\rho_2,\rho_3) =
\chi(\rho_2,\rho_3,\rho_1)\,,\qquad P^3=1\,,\qquad
 [{\cal H}_3,P]=[q_3,P]=0\,.
\end{equation}
To take the full advantage of the symmetry properties, we choose the
three independent functions $\Phi_{\la}(x)$ to be eigenfunctions of
the ``triality" operator $P$ and identify $\lambda$ as a corresponding
eigenvalue
\begin{equation}
\hat P\, \Phi_{\la}(x)=\lambda\Phi_{\la}(x)\,,\qquad
\lambda=1,\,  \e^{2\pi i/3},\,  \e^{-2\pi i/3}\,.
\label{trial}
\end{equation}
Under the cyclic permutations of holomorphic coordinates their anharmonic
ratio $x$ transforms as $x\rightarrow 1/(1-x) \rightarrow 1-1/x$
and the above condition, together with
$\hat P^2\Phi_{\la}=\lambda^2\Phi_{\la}$, translates into
\begin{eqnarray}
    \la \Phi_{\la}(x)&=&e^{-2\pi i h/3} ({-1/ x})^h \Phi_{\la}(1/(1-x)),
\nonumber \\[3mm]
   \la^2 \Phi_{\la}(x)&=& e^{-4\pi i h/3}
   ({-1/x})^h (x-1)^h \Phi_{\la}(1-1/x). \label{leqs}
\end{eqnarray}
Replacing $\Phi_{\la}(x)$ by its expression (\ref{filam}),
it is readily seen that these equations can be used to eliminate two of the initial
parameters , say $b_0/c_0$ and $c_1/c_0$ for each $\la$. With this choice
the remaining freedom reduces to a multiplicative normalization $c_0$.
We choose $c_0=1$ for each value of $\lambda$.
The relations (\ref{leqs}) should hold for any $x$ and
a particularly simple choice of $x$ will be discussed later.

A note on
the choice of unique branches in (\ref{leqs}) is in order. Even though the complete
wave function (\ref{odfun}) must be single-valued (see later), the eigenfunctions
$\Phi_{\lambda}(x)$ are in general multi-valued and the consistent choice of cuts
and phases is required. Eqs.(\ref{leqs}) correspond to the choice $|{\rm arg}(x)|<\pi,\;
{\rm arg}(-1/x)=\pi-{\rm arg}(x),\; |{\rm arg}(x-1)|<\pi$.

Identical considerations can be performed to construct the basis
functions $\bar\Phi_{\bar\la}(\bar x)$ in the  antiholomorphic
sector with $\bar\la$ being an eigenvalue of the operator of
cyclic permutations of the antiholomorphic coordinates $\bar P
\chi(\bar \rho_1,\bar \rho_2,\bar \rho_3)= \chi(\bar \rho_2,\bar
\rho_3,\bar \rho_1)$. One should notice that although the
operators $P$ and $\bar P$ act independently on the holomorphic
and antiholomorphic gluon coordinates, respectively, $[P,\bar
P]=0$, it is the operator $P \bar P$ that generates the cyclic
permutations of the gluons in the two-dimensional
$\vec\rho=(\rho,\bar\rho)-$plane. In addition, the relation $\bar
\rho=\rho^*$ implies that the definition of the cuts and phases of
the basis functions $\bar\Phi_{\bar\la}(\bar x)$ coincide with
those of $(\Phi_{\la}(x))^*$.

\section{Construction of the wave function}

Having defined the functions $\Phi_\lambda(x)$ and
$\bar\Phi_{\bar\lambda}(\bar x)$ we look for the wave function
$\Psi(x,\bar x)$ in the form (\ref{odfun}) with $\lambda$
and $\bar\lambda$ being cubic roots of unity and
$f_{\lambda\bar\lambda}$ being a $3\times 3$ matrix.
Thus constructed $\Psi(x,\bar x)$ is a simultaneous
eigenfunction of the integrals of motion $\hat q_k$ and $\bar {\hat q_k}$
for arbitrary $f_{\lambda\bar\lambda}$.
Let us now impose the condition that the odderon wave function
should be Bose symmetric single-valued function of gluon
coordinates on the 2-dim impact parameter $\vec\rho-$plane.

The Bose symmetry can be implemented in two steps. First,
requiring the symmetry under the cyclic permutations,
$P\bar P \Psi(x,\bar x)=\Psi(x,\bar x)$, and using (\ref{leqs})
we find the relation between holomorphic and antiholomorphic
trialities
\begin{equation}
    \la \bar{\la} = \exp{(-2\pi i(h-\bar{h})/3)}=\exp(-2\pi im/3) ,
\label{ll}
\end{equation}
with $m$ being the Lorentz spin of the state. Hence, for cyclic
symmetric wave function $\Psi(x,\bar x)$ the matrix
$f_{\lambda\bar\lambda}$ has only three nonvanishing entries
$f_\lambda\equiv f_{\lambda,\lambda^2\e^{-2\pi im/3}}$
%with $\lambda=1,\e^{\pm2\pi i/3}$
\begin{equation}
\Psi(x,\bar x)=\sum_{\lambda=1,\e^{\pm 2\pi i/3}
\atop \bar\lambda=\lambda^2\e^{-2\pi im/3}}
f_\lambda\,\Phi_\lambda(x)\bar\Phi_{\bar\lambda}(\bar x).
\label{WF}
\end{equation}
Finally, the full Bose symmetry is achieved through the
symmetrization $(\rho_1,\rho_2,\rho_3)\to (\rho_2,\rho_1,\rho_3)$,
or equivalently $x\to x/(x-1)$
\begin{equation}
\Psi_{\rm sym}(x,\bar x)=\Psi(x,\bar x)+\Psi\left({x\over x-1},
{\bar x\over\bar x-1} \right)\,.
\label{sym-WF}
\end{equation}

The expression (\ref{WF}) still contains three unknown parameters
$f_\lambda$. To fix them we impose the condition for $\Psi(x,\bar{x})$
to be a single-valued function around three singular points
$x=\bar x=0$, $x=\bar x=1$ and $x=\bar x=\infty$ corresponding
to the limit when any two of the reggeons are approaching each other. Due to
the cyclic symmetry of $\Psi(x,\bar x)$ it is sufficient to consider
only one of the singular points, say $x=\bar x=0$. To this end, we
substitute (\ref{filam}) into (\ref{WF}) and examine the asymptotic
behaviour of the functions (\ref{fiabc}) around $x=0$
\begin{eqnarray}
\phi^a(x)&=&x^{-h}\left(1+{i q_3\over(h-1)} x \log{x}+...\right),
\nonumber
\\
\phi^b(x)&=&x^0(1+ ... ),
\label{phi's}
\\[2mm]
\phi^c(x)&=&x^{-h+1}(1+...),
\nonumber
\end{eqnarray}
where $...$ denote terms suppressed by a power of $x$.
% and one has
Similar expressions hold in the antiholomorphic sector
\footnote{Incidentally Eq.(\ref{phi's}) establishes one to one
correspondence with the solutions of the third order differential
equation considered in \cite{PRL2}.}. It is easy to see, using
(\ref{phi's}), that for $q_3\neq 0$ and arbitrary complex (but not
integer) $h$ and $\bar h$ the only bilinear combination of the
functions $\phi^{(a,b,c)}(x)$ and $\bar\phi^{(a,b,c)}(\bar x)$, that is
well-defined around $x=\bar x=0$, has the following general form
\footnote{For $q_3=0$ the Bose symmetric wave function of the
three gluon compound state is given by the sum of the 2-gluon BFKL
wave functions, $\chi(\vec\rho_1,\vec\rho_2,\vec\rho_3)=
\chi_0(\vec\rho_1,\vec\rho_2)+\chi_0(\vec\rho_2,\vec\rho_3)
+\chi_0(\vec\rho_3,\vec\rho_1)$ with
$\chi_0(\vec\rho_1,\vec\rho_2)= \left({\rho_{12}/ \rho_{10}
\rho_{20}} \right)^{h} \left({\bar\rho_{12}/ \bar\rho_{10}
\bar\rho_{20}} \right)^{\bar h}$. However, this state is
degenerate and has a zero $SL(2)-$norm.}
\begin{eqnarray}
\Psi(x,\bar{x})&=&f_{cc}\phi^c(x)\bar{\phi}^c(\bar{x})+f_{bb}\phi^b(x)\bar{\phi}^b(\bar{x})
                             \nonumber \\
               &+&f_{ac}\phi^a(x)\bar{\phi}^c(\bar{x})+f_{ca}\phi^c(x)\bar{\phi}^a(\bar{x}),
                              \label{psifxy}
\end{eqnarray}
provided that $f_{ac}/f_{ca}=q_3(\bar h-1)/(\bar q_3(h-1))$. The
latter condition ensures that $\sim\ln(x)$ and $\sim\ln(\bar x)$ terms enter
with the same coefficients.
At the same time, replacing $\Phi(x)$ and $\bar\Phi(\bar x)$ in (\ref{WF}) by their
expressions (\ref{filam}) one obtains 9 different bilinear combinations
of the functions $\phi^{(a,b,c)}(x)$ and $\bar\phi^{(a,b,c)}(\bar x)$.
Matching them into (\ref{psifxy}) one finds
\begin{eqnarray}
f_{cc}=\sum_{\la} f_{\la}c_1(\la)\bar{c}_1(\bar\la), &&
f_{bb}=\sum_{\la} f_{\la}b_0(\la)\bar{b}_0(\bar\la), \nonumber \\
f_{ac}=\sum_{\la} f_{\la}        \bar{c}_1(\bar\la),&& \label{fxy}
f_{ca}=\sum_{\la} f_{\la}c_1(\la)              ,   \nonumber
\end{eqnarray}
provided that $\lambda$ and $\bar\lambda$ are related through
(\ref{ll}) and
\begin{equation}
{q_3\over h-1}\sum_{\la} f_{\la} \bar{c}_1(\bar{\la})=
{\bar{q}_3\over \bar{h}-1}\sum_{\la} f_{\la} c_1(\la)\,.
\end{equation}
We recall that throughout the paper we are using the normalization
$c_0(\lambda)=\bar c_0(\bar\lambda)=1$.
The requirement that the remaining 5 combinations do not appear in
(\ref{psifxy})
leads to the following conditions
\begin{equation}
   q_3\bar{q}_3  \sum_{\la} f_{\la} =
   \sum_{\la} f_{\la} \bar{b}_0(\bar{\la})=
   \sum_{\la} f_{\la} b_0(\la)=
   \sum_{\la} f_{\la} \bar{b}_0(\bar{\la})c_1(\la)=
   \sum_{\la} f_{\la} b_0(\la)\bar{c}_1(\bar{\la})=
   0
\label{cons}
\end{equation}
 The parameters $b_0(\lambda)$ and $c_1(\lambda)$ together
with their antiholomorphic counterparts depend on the
charges $q_3$ and $\bar q_3=q_3^*$ and, as we show in the next
section, can be uniquely defined from the triality conditions (\ref{leqs}).
The system of the equations (\ref{fxy}) and (\ref{cons}) on the coefficients
$f_{\la}$ is overcompleted and its consistency conditions provide the
quantization of $q_3$.

\section{Quantization conditions}

The odderon corresponds to the eigenstate $\Psi(x,\bar x)$ with
the maximal energy and is expected to have a zero Lorentz spin,
$m=0$, and pure imaginary value of the charge $q_3$. This allows us
to restrict further analysis to the lowest representation
\begin{equation}
h=\bar{h}=\frac12+i\nu\,,\qquad {\rm Re}\, q_3=0\,.
\label{h-sim}
\end{equation}

Let us now determine the coefficients $b_0(\lambda)$ and
$c_1(\lambda)$ {}from the triality conditions Eqs.(\ref{leqs}). To
this end we examine the asymptotic behavior of the both sides of
Eqs.(\ref{leqs}) in the limit $x\to 0$ and use the known
asymptotics of the conformal basis functions $\phi_{\al}$ around
$x=0,1$ and $\infty$
\begin{equation}
\phi_{\al}(x) \stackrel{x\to 0}{\sim} x^{\alpha-h}\,,\quad
              \stackrel{x\to 1}{\sim} \frac{\Gamma(2\alpha)\Gamma(h)}
              {\Gamma(\alpha+h)\Gamma(\alpha)}\,,\quad
              \stackrel{x\to\infty+i\epsilon}{\sim}\e^{i\pi(\alpha-h)}
               \frac{\Gamma(2\alpha)\Gamma(h)}
              {\Gamma(\alpha+h)\Gamma(\alpha)}\, .\quad
\end{equation}
Then, Eqs.(\ref{leqs}) are transformed into
\begin{eqnarray}
\lambda\e^{-i\pi h/3} &=&
\alpha(iq_3) + c_1(\lambda) \beta(iq_3) + b_0(\lambda) \gamma(iq_3),
\label{lineq}
\\
\lambda^2\e^{i\pi h/3} &=&
\alpha(-iq_3) + \left(\frac{\pi}{h-1}q_3 -c_1(\lambda) \right)
\beta(-iq_3) + b_0(\lambda) \gamma(-iq_3)\e^{i\pi h},
\nonumber
\end{eqnarray}
where the functions $\alpha$, $\beta$ and $\gamma$ depend only on
the charge $q_3$ and are defined through the series
\begin{eqnarray}
\alpha(iq_3)&=& \sum_{k=0}^\infty \left(
c_k + a_k \frac{d}{dk} \right)
\frac{\Gamma(h)\Gamma(2k)}
{\Gamma(k)\Gamma(k+h)}\bigg|_{c_0=1\,,\ c_1=0} \,,
\nonumber
\\[3mm]
\beta(iq_3)&=& \sum_{k=1}^\infty c_k \frac{\Gamma(h)\Gamma(2k)}
{\Gamma(k)\Gamma(k+h)}\bigg|_{c_0=0\,,\ c_1=1}\,,
\label{alpha}
\\[3mm]
\gamma(iq_3)&=&\sum_{k=0}^\infty b_k
\frac{\Gamma(h)\Gamma(2(k+h))}
{\Gamma(k+h)\Gamma(k+2h)}\bigg|_{b_0=1}\,, \nonumber
\end{eqnarray}
with the coefficients $(a_k,b_k,c_k)$ satisfying the recurrence
relations Eqs.~(\ref{rra})--(\ref{rrc}). The antiholomorphic
coefficients $\bar b_0$ and $\bar c_1$ obey similar equations and
 are related to their holomorphic counterpart as
\begin{equation}
\bar c_0=1\,,\qquad
\bar b_0(\bar\lambda)=b_0(\lambda) \e^{i\pi h}\,,\qquad
\bar c_1(\bar\lambda)=-c_1(\lambda)+\frac{\pi q_3}{h-1}.
\label{bar-bc}
\end{equation}
Using these relations one can resolve the system of equations
(\ref{fxy}) and (\ref{cons}) to obtain the explicit expressions for the coefficients
$f_\lambda$ as
\begin{equation}
f_{\lambda_i}=\varepsilon_{ijk}
\frac{c_1(\lambda_j)-c_1(\lambda_k)}{b_0(\lambda_i)},
\label{f's}
\end{equation}
with $i,j,k=1,2,3$ and $\lambda_k=\exp(2i \pi k/3)$. Then, the
quantization condition of $q_3$ takes a simple form (for $q_3\neq
0$)
\begin{equation}
%{\cal G}(q_3)\equiv
f_{\lambda_1}+f_{\lambda_2}+f_{\lambda_3}=0\,.
\label{QQ}
\end{equation}
We recall that the coefficients $c_1(\lambda)$ and $b_0(\lambda)$
entering (\ref{f's}) and (\ref{bar-bc}) are solutions of the system of linear
equations (\ref{lineq}). Solving (\ref{lineq}) it becomes straightforward
to express $f_\lambda$ in terms of the functions $\alpha(iq_3)$, $\beta(iq_3)$
and $\gamma(iq_3)$ defined in (\ref{alpha}). In this way one evaluates the
sum (\ref{QQ}) and identifies its zeros as quantized values of the
odderon charge $q_3$. The solutions to the quantization conditions are
shown in Fig.~1.

\begin{figure}[t]
%\vspace*{20mm}
%\hspace*{32mm}
\centerline{\epsfxsize9.0cm\epsfysize8.0cm\epsffile{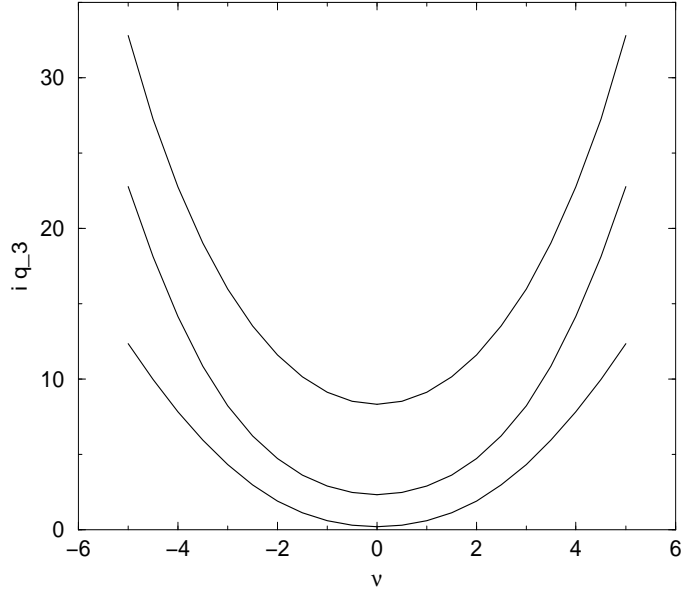}}
%\vspace*{-32mm}
\caption{The flow of the quantized values of the charge
$q_3=q_3(\nu,k)$ with the conformal weight $h=1/2+i\nu$ for the
lowest three trajectories $k=1,2,3$.}
\end{figure}

We find that in an agreement with the WKB analysis \cite{KorQ},
for any real $\nu$ defining the conformal weight of the state,
$(\ref{h-sim})$, the quantized $q_3$ takes a (infinite) series of
discrete values $q_3=q_3(\nu,k)$ that can be enumerated by a
positive integer $k$ with $k=1$ corresponding to the smallest
$|q_3|\neq 0$. Moreover, considering the dependence of $q_3$ on
real $\nu$ and integer positive $k$ we observe that $q_3(\nu,k)$
is a smooth function of $\nu$ for given $k$ and, therefore, the
quantized values of $q_3$ form a family of curves parameterized by an
integer $k$. The lowest three curves corresponding to $k=1\,,2\,,$
and $3$ are shown in Fig.~1.

\section{Special case: $h=1/2$}

%\subsection{The eigenvalues}
\underline{{\em 1. The eigenvalues.}}
Additional important simplification occurs for the smallest
absolute value of the conformal weight, $h=\bar h=\frac12$, or
equivalently $\nu=0$. Using the fact that $\alpha$, $\beta$ and
$\gamma$ are real functions of $iq_3$ one finds that in this case
the quantization condition (\ref{QQ}) can be rewritten in one of
the following {\it equivalent\/} forms
\begin{eqnarray}
\gamma(iq_3)&=&0\,,\nonumber \\
 2i\pi q_3 \beta^2(iq_3)&=&\beta(-iq_3), \nonumber \\
 \alpha(iq_3)\beta(-iq_3)&=&-\alpha(-iq_3)\beta(iq_3).  \label{rem}
\end{eqnarray}
Considering the first relation and using the definition (\ref{alpha})
%identity
%$\Gamma(1/2)\Gamma(2k+1)/\Gamma(k+1/2)\Gamma(k+1)=4^k$, one
one obtains the quantization condition for $q_3$ at $h=\bar h=\frac12$
in the form
\begin{equation}
%{\cal G}(q_3) \sim
\gamma(iq_3) = \sum_{k=0}^\infty 4^k b_k(q_3) = 0\,.
\end{equation}
That is, the eigenvalues of $q_3$ are the zeroes of the generating function
of the $b_k$ coefficients considered as a function of $q_3$.
%The coefficients $b_k$ obey the three-term recurrence relations
%(\ref{rrb}).
 %Then,
 Introducing new coefficients
$\hat{b}_k=(-4)^k k(k^2-1/4)b_k$, such that
$4 i q_3 b_k =(-4)^{-k}\left(\hat b_{k+1}-\hat b_{k-1}\right)$, one gets
\begin{equation}
\gamma(iq_3) = \hat b_{2\infty+1}(q_3) - \hat b_{2\infty}(q_3) = 0\, .
\label{infinity}
\end{equation}
Here $\hat b_{2\infty+1}$ and $\hat b_{2\infty}$ are limiting
values of the coefficients $\hat b_k$ for odd and even $k$,
respectively. To find their values one considers the three term
recurrence relations for $\hat b_k$
\begin{equation}
4iq_3 \hat b_k = k\left(k^2-\frac14\right)
\left[\hat b_{k+1}-\hat b_{k-1}\right]
\,,\qquad
\hat b_{0}=0\,,\quad \hat b_1=1\,.
\label{hat-b}
\end{equation}
Their solution can be found using a $2\times 2$ transfer matrix as
\begin{equation}
\left( \begin{array}{c}
            \hat{b}_{n+1} \\
            \hat{b}_{n}
       \end{array} \right) = \prod_{k=1}^n
\left( \begin{array}{cc}
            v_k & 1 \\
            1    & 0
       \end{array} \right) \left( \begin{array}{c}
                                   1 \\
                                   0
                                     \end{array} \right),
\qquad v_k={4 i q_3\over k(k^2-1/4)}.
\end{equation}
Then, it is straightforward to show that $\hat b_{2\infty+1}$ and
$\hat b_{2\infty}$ are the smooth, real functions of $iq_3$ and
solutions to (\ref{infinity}) define an infinite set of
discrete real positive values of $iq_3$.
At small $iq_3$ one applies the recurrence relations (\ref{hat-b})
to obtain the first few terms of the expansion of $\gamma(q)$
\begin{eqnarray}
\gamma(q)&=&
%1- 8\ln 2\, q+ \xi q^2 + {\cal O}(q^3)
1
- 8\ln 2\,q
%- 5.545177\,q
+ 3.4066793\,{q}^{2}
- 0.6271540\,{q}^{3}
+ 0.0494534\,{q}^{4},
\nonumber
\\
&-&
 0.0020214\,{q}^{5}
+ 0.0000483\,{q}^{6}
-0.0000007\,{q}^{7}
+ {\cal O}(q^8)
\label{smq}
\end{eqnarray}
%and $\xi=4\sum_{n=1}^\infty
%\frac{\Psi(2n+1/2)-\Psi(n+1/2)}{n(n^2-1/16)} =3.40668...$
where the expansion coefficients starting from ${\cal
O}(q^2)-$term can be expressed in terms of the hypergeometric series.

The expansion (\ref{smq}) can be used to obtain the lowest values
of quantized $q_3$. In Table 1 we quote the first three
eigenvalues. High precision solutions of Eq.(\ref{QQ}) are
displayed in the first column \footnote{In order to reach quoted
precision it is numerically advantageous to solve Eqs.(\ref{leqs})
at the Lagrange point $x_L=\exp(i\pi/3)$ point (c.f. Section 6)
where the convergence of $k-$ summation is exponential.}. These
numbers fully agree with the results obtained in \cite{PRL2} and
subsequent more precise estimates \cite{Brau}.  The first three
zeroes of the small $q_3$ expansion, Eq.(\ref{smq}), are quoted in
the second column. They agree extremely well with the precise
results for the lowest state. Even for the third state the small
$q_3$ approximation works qualitatively.

%For higher values of $k$ the charge
%scales as $q_3^{(k)}\sim k^3/\sqrt{27}$ in agreement with the
%WKB asymptotics \cite{KorWh}.

  \begin{table}
  \begin{center}
   \begin{tabular}{|c|cc|c|} \hline\hline
   {\it No.} & \multicolumn{2}{c|}{ $iq_3$ }& $iq_3^{\rm as}$
                                      \\
   \hline
%  0 & \multicolumn{2}{c|}{$0$}
%                                     \\
%
  1 & \multicolumn{2}{c|}{$0.20525750608820 $} & 0.2052575
                                      \\
  2 & \multicolumn{2}{c|}{$2.34392106326404 $} & 2.343918
                                     \\
  3 & \multicolumn{2}{c|}{$8.32634590161324 $} & 8.307685 \\
   \hline\hline
   \end{tabular}
  \end{center}
\caption{Quantization of $q_3$.}
   \end{table}

%\subsection{The wave function}
\noindent \underline{{\em 2. The wave function.}}
For $h=\bar h=1/2$ additional simplification occurs for the
odderon wave function as well. Firstly, the basis of
$SL(2,\mathbb{C})$ harmonics, (\ref{harmo}), reduces to a simple
power
\begin{equation}
\phi_{\al}(x) = (4u)^{\al-1/2}\,,\qquad
   u={1-\sqrt{1-x} \over 1+\sqrt{1-x}},  \label{uvar}
\end{equation}
which suggests an important role of the new variable $u$ and its antiholomorphic counterpart
$\bar u$ defined in a similar way. In new variables the
completely symmetric odderon wave function reads, $\varphi={\rm arg\/}(u)$
\begin{eqnarray}
\Psi_{\rm sym}(x,\bar x)&=&\sum_{m\ge 0, m={\rm even}}^{\infty}
2\cos(m\varphi)S_m(|u|), \label{urep}\\
S_m(|u|)&=&\sum_{r\ge|m|,r={\rm even}}^{\infty}
C_{mr}|u|^{r-1}+B_{mr}|u|^r
         +2A_{mr}|u|^{r-1}\log{(4|u|)}, \nonumber
\end{eqnarray}
where the coefficients $A\,,B$ and $C$ are given by
\begin{eqnarray}
&&A_{mr}=4^{r-1}f_{ac}\bar{c}^{(c)}_k=4^{r-1}f_{ca}c^{(c)}_k,
% \nonumber\\ &&
B_{mr}=4^r f_{bb}b_k\bar{b}_n,  \nonumber\\
&&C_{mr}=4^{r-1}(f_{cc}c^{(c)}_k\bar{c}^{(c)}_n+f_{ac}c^{(a)}_k\bar{c}^{(c)}_n
                +f_{ca}c^{(c)}_k\bar{c}^{(a)}_n),  \label{ABC}
%\\ && k=\frac12(r+m),\;\;n=\frac12(r-m)  \nonumber
\end{eqnarray}
$k=(r+m)/2$, $n=(r-m)/2$
and $c^{(a,c)}_k$ denote the expansion coefficients entering the
expansion of the functions $\phi^a(x)$ and $\phi^c(x)$,
respectively, defined in Eq.(\ref{fiabc}).

This representation is convergent in the unit circle $|u|\leq 1$ onto which
the cut complex plane of $x$ is mapped under (\ref{uvar}) with both edges of the cut
 $1<x<\infty$ transformed onto the circumference $|u|=1$. It is worth noting that
the function $S_m(\rho)$ satisfies the relations $S_{odd}(1)=0$ and $d S_{even}(1)/d\rho=0$,
which guarantee smoothness of $\Psi_{\rm sym}$
across the cut $1<x<\infty$.

Since the basic conditions imposed on the compound eigenfunction (\ref{WF})
 (uniqueness and Bose symmetry),
as well as the resulting structure (\ref{psifxy}), are the same as
in Ref.\cite{PRL2} it is natural that the resulting wave functions
are identical with those constructed there. We have found a complete
numerical agreement for the three lowest states. However the present
approach is advantageous in the two respects. First, it gives much
simpler (although equivalent) quantization of $q_3$, and second,
the representation (\ref{fiabc}) is {\em convergent} in the whole
transverse plane while earlier expression required the analytical
continuation between the three domains of $x$.

\section{Observables and sum rules}

Having constructed the eigenvalues and the eigenfunctions we turn now to the
observables. A large class of operators, relevant to the odderon problem,
consists of the  $SL(2,\mathbb{C})$ invariant operators symmetric under the cyclic
permutations and having a pairwise structure, 
say ${\cal O}=f(L_{12}^2)+f(L_{23}^2)+f(L_{31}^2$).
Their eigenvalues can be readily extracted using the expansion (\ref{inal}).
%In fact this
%property was the original motivation for our choice of basis (\ref{harmo}).
%Consider the symmetric two body operator $\hat{O}=f(L_{12}^2)+f(L_{23}^2)+f(L_{31}^2)$.
Even though the three terms entering ${\cal O}$ do not commute with each other,
the symmetry of the wave function (\ref{trial}) under $P$ allows us to calculate the
action of ${\cal O}$ on the states $\Phi_{\la}(x)$. Using the
identities $f(L_{23}^2)=P f(L_{12}^2)P^2$ and $f(L_{31}^2)=P^2
f(L_{12}^2)P$ and replacing $\Phi_{\la}(x)$ by its expansion (\ref{inal})
we get
\begin{equation}
{\cal O} \Phi_{\la}(x)=(1+\la^2 P  + \la P^2)f(L_{12}^2)\Phi_{\la}(x)
= \int_\Gamma \frac{d\al}{2\pi i}\ C(\al)f(\al(\al-1))
\varphi_\alpha(x;\lambda),
\label{osr}
\end{equation}
where
\begin{equation}
\varphi_\alpha(x;\lambda)=\phi_{\al}(x)
   +\la^2 \e^{-2\pi i h/3} (-1/x)^{h}\phi_{\al}\left(\frac{1}{1-x}\right)
  + \la \e^{-4\pi i
  h/3}(-1/x)^{h}(x-1)^h\phi_{\al}\left(1-\frac1{x}\right)\,.
\end{equation}
The integration in (\ref{osr}) reduces to the discrete sum over
poles of the coefficient functions at $\alpha=k,\; k+h$.
By an appropriate symmetric choice of $x$ this formula can be yet simplified. At
the Lagrange point
\footnote{The Lagrange point corresponds to three bodies (gluons) forming an equilateral
triangle - a configuration well known in astronomy.},
$x_L=\exp(i\pi/3)$, such that $x_L=1/(1-x_L)=1-1/x_L$ one finds that
\begin{equation}
\varphi_\alpha(x_L;\lambda)=\phi_{\al}(x_L)(1+\la+\la^2) = 3
\phi_\al(x_L),
\end{equation}
for $\lambda=1$ and $\varphi_\alpha(x_L;\lambda)=0$ for $\lambda=\exp(\pm
2i\pi/3)$. Therefore,
\begin{equation}
  {\cal O}\,\Phi_\la(x_L)=3\delta_{\la,1} f(L_{12}^2)\,\Phi_\la(x_L)
=3\delta_{\la,1}\int_\Gamma \frac{d\al}{2\pi i}\ C(\al)f(\al(\al-1))
\phi_\alpha(x_L)\,. \label{lsr}
\end{equation}
%i.e. it is sufficient to sum only the first term in (\ref{osr}).
Another advantage of the Lagrange point is that for $h=\bar h=1/2$
% are also very usefull for practical purposes.
the phase of the harmonics $\phi_{\al}(x_L)$ given by Eq.(\ref{uvar}) 
is very simple and
the absolute value of $\phi_{\al}(x_L)$ is small, which makes the
sum over residues in (\ref{lsr}) exponentially convergent.

If $\Phi_\la(x)$ is the eigenfunction of ${\cal O}$, then the
relation (\ref{lsr}) allows for calculating the corresponding
eigenvalue. On the other hand, comparing the $x$-dependence of the
both sides of the general formula (\ref{osr}) one could verify
whether $\Phi_{\la}(x)$ is the {\em eigenfunction} of $\cal O$.
In particular this is the case for ${\cal
O}=L^2=L^2_{12}+L^2_{23}+L^2_{31}$. Moreover, since by the
construction $\Phi_\la(x)$ diagonalizes the Casimir operator $L^2$,
and the resulting eigenvalue is known to be $h(h-1)$, the above
equations provide useful sum rules to test our construction of the
wave function. We verified that for $h=\bar h=1/2$ these sum rules
are satisfied to high accuracy independently of $x$.
%, so indeed our
%construction provides appropriate eigenfunctions of $L^2$.
%On the other hand, the $x$-dependence in the general formula (\ref{osr})
% allows to verify if $\Phi_{\la}(x)$ is the {\em eigenfunction} of $\hat{O}$.
%Namely it is sufficient to check if $\hat{O}\Phi_{\la}(x)/\Phi_{\la}(x)$
%does not depend on $x$ \footnote{Also the equivalent differential conditions can be
%formulated.}.
%   We have tested Eqs.(\ref{osr},\ref{lsr}) numerically for the
%total angular momentum $L^2=L^2_{12}+L^2_{23}+L^2_{31}$, for $h=1/2$.
%Since in this case the resulting eigenvalue is known to be $h(h-1)$,
% above equations provide useful sum rules to test our construction of the wave
%function. Both checks, i.e. the eigenvalue and the eigenfunction tests worked
%satisfactorily to high accuracy.

Analogous sum rules hold for the complete eigenfunction
(\ref{psifxy}). However, while the sum rules (\ref{osr}) for
$\Phi_\la(x)$ are satisfied for arbitrary $q_3$, similar relations
for the eigenfunction $\Psi(x,\bx)$ \footnote{With the analog of
the rhs of Eq.(\ref{osr})
 given essentially by Eq.(\ref{urep}) with obvious modifications.}
\begin{equation}
(L^2+\bar{L}^2)\Psi(x,\bar{x})/\Psi(x,\bar{x})=2 \,{\rm Re}\, h(h-1)\,, \label{hhsr}
\end{equation}
are satisfied only for the quantized values of $q_3$, since in
arriving at Eqs.(\ref{psifxy}) and (\ref{urep}) we have used the 
quantization conditions (\ref{fxy}) and (\ref{cons}).
In fact, a high numerical precision quoted in the first column of Table 1 was required
to satisfactorily reproduce rhs of (\ref{hhsr}) at $h=\bar h=1/2$ for the first three
states in a large range of $x$.

It becomes straightforward to apply the above method to calculate
the eigenvalues of the Hamiltonian, (\ref{H-QCD}). Using the
cyclic symmetry, its action on the wave function $\Phi_\lambda(x)$
can be reduced at the Lagrange point to that of $H_{12}+\bar
H_{12}$ only. Moreover, because of (\ref{hik}), Eq.(\ref{osr})
with $f(\alpha)=\psi(\al)+\psi(1-\al)-2 \psi(1)$ would apply.
Similarly one could calculate the action of the Hamiltonian
$\mathbb{H}_3$ on the wave function (\ref{WF}). We note, however,
that the two body holomorphic energy, $f(\alpha)$, has additional
poles at integer values of the conformal weight $\alpha$, which
could interfere with the singularities of the coefficient function
in (\ref{lsr}). Although these poles cancel out in the sum of the
holomorphic and antiholomorphic energies $f(\alpha)+f(\bar\alpha)$,
provided that $\bar\alpha$ and $\alpha$ are taking values in the
principal series representation of the $SL(2,\mathbb{C})$, one has
to introduce an additional prescription to separate them from the
singularities of the coefficient function $C(\alpha)$.

\section{Conclusions}

New representation of the odderon wave function was derived, which
is convergent in the whole impact parameter plane, and provides the
analytic form of the quantization condition for the integral of motion $q_3$.
Solving these quantization conditions we have found that the quantized values of $q_3$
form smooth trajectories parametrized by an integer.
Our results are in agreement with the %original
findings of Ref.\cite{PRL2} and earlier WKB expressions \cite{KorQ}.
A new quantum number (triality) was
identified independently in each sector. This, together with the choice of the
conformal basis, allows for simple calculation of the eigenvalues
of a wide class of operators. Angular momentum sum rules were derived and shown
to provide a useful test of the whole scheme. The odderon Hamiltonian also belongs
to the above class, however its action on the conformal basis requires additional
studies.

\vspace*{1cm}
%The angular momentum basis was independently introduced by Lipatov and discussed
%in general terms in \cite{LIPRP}.
We thank R. A. Janik for numerous discussions. J. W. thanks L. Lipatov for the
 discussion. J.W. also thanks the CNRS for the financial support.
This work is  partially supported by the Polish Committee for Scientific 
Research under grants no. PB 2 P03B 044 12 and PB 2 P03B 010 15.

\end{document}